\documentclass[12pt]{iopart}
\usepackage{graphicx,color,amssymb,amsfonts,hyperref}
\usepackage[frenchb,english]{babel}

\newcommand{\mP}{{\mathcal P}}

\newcommand{\mB}{{\mathcal B}}
\newcommand{\mV}{{\mathcal V}}

\newcommand{\mR}{{\mathcal R}}

\newcommand{\cv}{{\rm Conv}}

\newcommand{\spn}{{\rm Span}}

\newcommand{\ba}{\mathbf  a}
\newcommand{\bb}{{\bf b}}
\newcommand{\bk}{{\bf c}}

\renewcommand{\bs}{{\bf s}}
\newcommand{\bt}{{\bf t}}

\definecolor{nblue}{rgb}{0.3,0.3,1.0}
\definecolor{ngreen}{rgb}{0.2,0.7,0.2}
\definecolor{nred}{rgb}{0.9,0.1,0}
\definecolor{npurple}{rgb}{0.8,0.2,0.8}
\definecolor{golden}{rgb}{0.8,0.6,0.1}
\definecolor{nsilver}{rgb}{0.3,0.4,0.5}
\definecolor{nbrown}{rgb}{0.8,0.4,0.15}
\definecolor{nrose}{rgb}{0.7,0,0.35}
\definecolor{nviol}{rgb}{0.5,0,1.0}
\definecolor{nazur}{rgb}{0,0.35,0.7}
\definecolor{nchart}{rgb}{0.2,0.4,0}

\begin{document}

\title{Bell's inequality and extremal nonlocal box from Hardy's test for nonlocality} 
\author{Sixia Yu}
\address{Centre for Quantum Technologies, National University of Singapore, 2 Science Drive 3, Singapore 117542}

\begin{abstract} Bell showed 50 years ago that quantum theory is nonlocal via his celebrated inequalities, turning the issue of quantum nonlocality from a matter of taste into a matter of test. Years later, Hardy proposed a test for nonlocality without inequality, which is a kind of ``something-versus-nothing" argument. Hardy's test for $n$ particles induces an $n$-partite Bell's inequality with two dichotomic local measurements for each observer, which has been shown to be violated by all entangled pure states. Our first result is to show that the Bell-Hardy inequality arising form Hardy's nonlocality test is tight for an arbitrary number of parties, i.e., it defines a facet of the Bell polytope in the given scenario. On the other hand quantum theory is not that nonlocal since it forbids signaling and even not as nonlocal as allowed by non-signaling conditions, i.e., quantum mechanical predictions form a strict subset of the so called non-signaling polytope. In the scenario of each observer measuring two dichotomic observables, Fritz established a duality between the Bell polytope and the non-signaling polytope: tight Bell's inequalities, the facets of the Bell polytope, are in a one-to-one correspondence with extremal non-signaling boxes, the vertices of the non-signaling polytope. Our second result is to provide an alternative and more direct formula for this duality. As an example, the tight Bell-Hardy inequality gives rise to an extremal non-signaling box that serves as a natural multipartite generalization of Popescu-Rohrlich box.
\\

\noindent This article is part of a special issue of {\it Journal of Physics A: Mathematical and Theoretical} devoted to `50 years of Bell's theorem'.
\\

\end{abstract} 

\maketitle

\section{Introduction}

Fifty years ago, Bell \cite{Bel64,Bel66} proved that quantum theory is nonlocal by showing that quantum mechanical predictions on a finite set of correlations of local measurements cannot be reproduced by any local realistic model via his celebrated inequality, Bell's inequality. Ever since different types of Bell's inequalities in various scenarios have been proposed and the violations to Bell's inequality have been verified in many experiment situations, finding numerous applications in various quantum informational processes \cite{rmp} (and references therein). Twenty-five years ago Werner \cite{werner} introduced the exact definition of quantum entanglement and proved that entanglement is necessary but not sufficient for the the violations of Bell's inequality. However, for pure states these two fundamental features of quantum theory, quantum nonlocality and entanglement, become equivalent, known as Gisin's theorem \cite{gisin}. 

The equivalence between the quantum nonlocality and entanglement was established  by Gisin \cite{gisin} and Gisin and Peres \cite{gisin2} in the case of two particles by proving that all entangled two-particle pure states violate a single Bell's inequality, due to Clauser, Horne, Shimony, and Holt (CHSH) \cite{chsh}. In the case of three qubits Gisin's theorem was proved by Chen {\it et al}. \cite{chen} numerically and by Choudhary {\it et al}. \cite{chou}. In the most general case the equivalence was established by Popscu and Rohrlich (PR) \cite{pr} using a set of Bell-CHSH inequalities and more recently by Yu {\it et al}. \cite{yu.gisin} using a single Bell's inequality arising from Hardy's test for nonlocality, which is referred to as Bell-Hardy inequality. Years after Bell's theorem, Greenberger, Horne, and Zeilinger (GHZ) \cite{ghz} provided a compelling  argument, which is  elegantly formulated by Mermin \cite{mermin.ghz}, for the quantum nonlocality of some special states, e.g., GHZ states as well as some graph states \cite{tang.yu.oh}, without inequality. This kind of ``all-versus-nothing" argument is applicable only to three or more particles. Soon after, Hardy \cite{hardy1,hardy2} provided a nonlocality test for two particles without inequality, which can be regarded as a kind of a ``something-versus-nothing" argument.

Mermin \cite{mermin.hd} formulated a Bell inequality from two-particle Hardy's test, which is equivalent to the Bell-CHSH inequality and is generalized by Cereceda \cite{cereceda} to the case of $n$ particles. From the point view of the equivalence between nonlocality and entanglement the Bell-Hardy inequality is the most natural generalization of the two-particle Bell-CHSH inequality into many particles. Moreover it is well known that the Bell-CHSH inequality is tight, i.e., a facet of the Bell polytope in the scenario of two observers each of which measures two dichotomic observables. Our first result is to prove that the Bell-Hardy inequality for an arbitrary number of observers is also tight, providing another justification for regarding the Bell-Hardy inequality as a most natural generalization of the Bell-CHSH inequality. As is well known tight Bell's inequalities  completely delineate the correlations that are explainable by local realistic theory  in a given scenario and thus serve as optimal witnesses of nonlocal correlations.

On the other hand quantum theory is not that nonlocal since it forbids signaling, i.e., sending information via local choices of different measurement settings is impossible. There are many efforts to understand why this is the case for quantum theory by introducing some additional constraints other than non-signaling, such as information causality \cite{inf.c} and exclusion principle \cite{ge} or local orthogonality \cite{exc}. In a given scenario, as all local correlations form the so-called Bell polytope,  all non-signaling correlations for the so-called non-signaling polytope, which include the quantum theory as a strict subset because there are non-signaling correlations, e.g., the PR box \cite{pr.box}, whose correlations are too strong to be reproduced even in quantum theory. An important progress of understanding non-signaling polytope is made by Fritz \cite{fritz} by establishing a one-to-one correspondence of the extremal points of non-signaling polytope and the facets of Bell polytope in the special scenario of each observer measuring two dichotomic observables. In this case every tight Bell's inequality gives rise to an extremal non-signaling box and vice versa. 
For example the extremal non-signaling box corresponding to the Bell-CHSH inequality is exactly the well-known PR box. Our second result is to provide an alternative proof for Fritz's duality with a more direct formula. As an example we derive the extremal nonlocal box that corresponds to the tight Bell-Hardy inequality.

\section{Bell polytope}

In an $(n,s,d)$ scenario there are $n$ observes that are mutually space-like separated and each observer measures $s$ observables with $d$ outcomes. Most generally each observer may perform a different number of measurements and each measurement may have a different number of outcomes. 
In what follows we consider only the case $s=d=2$, i.e., there are $n$ observers each of which perfuming locally two dichotomic measurements. 

Let $n$ observers be labeled with an index set $I=\{1,2,\ldots,n\}$ and each observer, e.g., observer $k\in I$,  measures two dichotomic observables $A_{k|0}$ and $A_{k|1}$ with two outcomes labeled with $a_{k|0},a_{k|1}=0,1$ respectively. We denote by  $\bs=[s_k]_{k\in I}$ the setting vector with each component $s_k=0,1$ labeling two observables that are measured by the $k$-th observer. For a given measurement setting $\bs$ we denote by $\ba_\bs=[a_{k|s_k}]_{k\in I}$ the corresponding outcome vector  with each component $a_{k|s_k}=0,1$ labeling two outcomes of measuring the observable $A_{k|s_k}$ with $k\in I$. Both two  vectors $\bs$ and $\ba_\bs$ are $n$-dimensional binary vectors and we denote by $2^I$ the set of all $n$-dimensional binary vectors. Slightly abusing the notations we shall use a binary vector $\ba\in 2^I$ to denote also a subset of $I$, i.e., its support $\{k\in I|a_k\not=0\}$,  and by $|\ba|$ the number of nonzero components of a vector $\ba$. For later use we shall build from  two  vectors $\ba,\bb\in 2^I$ a third $n$-dimensional binary vector $\ba\wedge\bb=[a_kb_k]_{k\in I}$. 

There are in total $4^n$ events labeled with the setting vector $\bs$ and the corresponding outcome vector $\ba_\bs$ and each event is associated with a probability $P(\ba_\bs|\bs)$ in a probabilistic theory, such as quantum theory or local hidden variable model.
In a local realistic model, for a given hidden variable $\lambda$ distributed according to $\varrho_\lambda$, we denote by $a_{k|s_k}^\lambda\in \{0,1\}$ the outcome of observer $k$ measuring observable $A_{k|s_k}$ for $k\in I$. The joint probability of the event that observer $k$ measures observable $s_k$ obtaining outcome $a_{k|s_k}$ with $k\in I$ then assumes the following local form
\begin{equation}\label{P}
P(\ba_\bs|\bs)=\int d\lambda \ \varrho_\lambda \prod_{k\in I}\delta(a_{k|s_k}^\lambda, a_{k|s_k}).
\end{equation}
Every set of $4^n$ probabilities $\{P(\ba_\bs|\bs)\mid\bs,\ba_\bs\in 2^I\}$ of above form, also referred to as a local box here, can be viewed as a vector in a $4^n$ dimensional  space $\mR$, called as box space. The box space $\mR$ can be conveniently taken as a real subspace of the Hilbert space of $2n$ qubits and we shall label those $n$ pairs of qubits by the index set $I$ with $\{|\bs,\ba_\bs\rangle=|\bs\rangle_I\otimes|\ba_\bs\rangle_I\mid\bs,\ba_\bs\in 2^I\}$ being an arbitrary basis. In the box space $\mR$ each local box $\{P(\ba_\bs|\bs)\mid\bs,\ba_\bs\in 2^I\}$ is represented by a vector 
\begin{equation}
|P\rangle=\sum_{\bs\in 2^I}\sum_{\ba_\bs\in 2^I}P(\ba_\bs|\bs)|\bs,\ba_\bs\rangle
=\int d\lambda\ \varrho_\lambda\bigotimes_{k\in I}\sum_{s_k=0}^1|s_k\rangle_k\otimes|a_{k|s_k}^\lambda\rangle_k
\end{equation}
that is a convex combination of $4^n$ vectors in
\begin{equation}\label{ep}
P_n=\{|\ba;\bb\rangle=\otimes_{k\in I}\left(|0\rangle_k\otimes|a_k\rangle_k+|1\rangle_k\otimes|b\rangle_k\right)\mid \ba,\bb\in 2^I\}.
\end{equation} 
On the other hand every possible convex combination of vectors in $P_n$ also corresponds to a possible set of statistical predictions by the local realistic model in the given scenario. 
Thus all possible statistical predictions of a local realistic model are identical with the vectors in the so-called Bell polytope $\mB_n=\cv(P_n)$ formed by all possible convex combinations of vectors in $P_n$. Those vectors in $P_n$ are vertices, or extremal vectors, of the Bell polytope.  Because of the following identities
\begin{equation}\label{id}
|0;0\rangle_k+|1;1\rangle_k=|0;1\rangle_k+|1;0\rangle_k\quad (\forall k\in I),
\end{equation}
where we have denoted
$
|a;b\rangle=|0\rangle\otimes|a\rangle+|1\rangle\otimes|b\rangle, 
$ there are at most $3^n$ linearly independent vectors in $P_n$. By noting the direct product structure of the vectors in $P_n$, it is straightforward to compute the Gramm matrix of all $3^n$ vectors in $P_n$ which can be readily found to be nonsingular. Thus the dimension can be exactly determined to be $3^n$.  Therefore  the linear span $\mP_n=\spn(P_n)$ of  $P_n$, i.e., the set of all vectors that are linear combinations of vectors in $P_n$, is of dimension $\dim \mP_n=3^n$. As a result the Bell polytope $\mB_n$ corresponding to $(n,2,2)$ scenario is of dimension $3^n-1$, i.e., there are $3^n-1$ independent parameters in a local box. Properties of Bell polytope in general scenario can be found in Ref.\cite{pironio}. 
Those independent parameters in a local box can be taken as the following $3^n-1$ correlations
\begin{equation}
A_\bs^\bk=\sum_{a_\bs\in 2^I}(-1)^{\ba_\bs\cdot \bk}P(\ba_\bs|\bs)\quad (\bs\subseteq c),
\end{equation}
with properties $A^{\bf 0}_{\bf 0}=1$ and $A^\bk_\bs=A^\bk_{\bs\wedge\bk}$ for a general $\bs$.

In terms of Bell polytope, a Bell's inequality is nothing else than a hyper plane that divides the $4^n$ dimensional box space $\mR$ into two parts with the Bell polytope being completely contained in one part, e.g., $\langle B|P\rangle\ge0$ for all $|P\rangle\in \mB_n$ for a fixed vector $|B\rangle\in\mR$. A facet of the Bell polytope is a hyper plane containing exactly $3^n-1$ linearly independent extremal vectors in $P_n$ and naturally defines a Bell's inequality, which is referred to as a tight Bell's inequality. Precisely, a Bell's inequality $\langle B|P\rangle\ge0$ in the scenario $(n,2,2)$ is called as tight if and only if there are $3^n-1$ independent extremal vectors $|\ba;\bb\rangle\in P_n$ such that $\langle B|\ba;\bb\rangle=0$. 
The following Lemma is about two standard forms of a Bell's inequality.

{\bf Lemma } Every Bell's inequality can be cast into a form involving probabilities and a from involving correaltions
\begin{equation}\label{bi}
\langle B|P\rangle=\sum_{\bs,\ba_\bs\in 2^I}B(\ba_\bs,{\bs})P(\ba_\bs|{\bs})=\sum_{\bk\in 2^I}\sum_{\bs\subseteq\bk}B^\bk_\bs A^\bk_\bs\ge 0,
\end{equation}
where  two sets of coefficients  are related to each other via
\begin{equation}\label{bk}
 B_\bs^\bk=\frac1{2^{|\bk|}}\sum_{a_\bs\in 2^I}(-1)^{\ba_\bs\cdot \bk}B(\ba_\bs,\bs)
\end{equation}
with real coefficients $B(\ba_\bs,{\bs})$ satisfying
\begin{equation}\label{cd}
\sum_{\bs,\ba_\bs\in 2^I}B(\ba_\bs,\bs)=1,\quad \sum_{a_{k|s_k}=0}^1B(\ba_\bs,\bs)\ \mbox{is independent of}\ s_k \quad (\forall k\in I).
\end{equation}

{\bf Proof.} Being determined by a hyper plane in $\mR$, every Bell's inequality assumes the form $\langle B|P\rangle\ge0 $ for some fixed vector $|B\rangle\in \mR$  and we have only to prove that its coefficients $B(\ba_\bs,{\bs})=\langle B|\bs\rangle_I\otimes|\ba_\bs\rangle_I$ can be chosen to satisfy conditions Eq.(\ref{cd}). For an arbitrary extremal vector $|\bb_{\bf 0};\bb_{\bf 1}\rangle\in P_n$ of the Bell polytope Bell's inequality must hold 
\begin{equation}\label{bs}
\langle B|\bb_0;\bb_1\rangle=\sum_{\bs\in 2^I}B(\bb_\bs,\bs)\ge0.\end{equation}
Since $\{|P\rangle|\langle B|P\rangle=0\}$ defines a hyper plane in the box space $\mR$ with one dimension less than the Bell polytope, there is at least one extremal vector $|\bb_{\bf 0};\bb_{\bf 1}\rangle\in P_n$ of the Bell polytope such that $\langle B|\bb_0;\bb_1\rangle\not =0$.  That is to say  the inequality Eq.(\ref{bs}) cannot be attained for all $\bb_\bs$ from which it follows  
\begin{equation*}
\langle B|\theta\rangle=\sum_{\bs,\bb_\bs\in 2^I}B(\bb_\bs,\bs)>0,\quad |\theta\rangle=\sum_{\bs\in 2^I}|\bs\rangle_I\otimes\sum_{\ba\in 2^I}|\ba\rangle_I.
\end{equation*}
Thus the first condition in Eq.(\ref{cd}) can be satisfied by dividing both sides of Bell's inequality by a positive number $\langle B|\theta\rangle$.
Secondly,
we can rewrite Bell's inequality for probabilities in Eq.(\ref{bi}) as
\begin{equation}
0\le\sum_{\bk,\bs\in 2^I}\frac{B^\bk_\bs A^\bk_\bs}{2^{n-|\bk|}}=\sum_{\bk\in 2^I}\sum_{\bs\subseteq\bk}\frac{A_\bs^\bk}{2^{n-|\bk|}}\sum_{\bt\in 2^I}B_{\bt}^\bk \delta_{\bt\wedge\bk,\bs\wedge\bk}
:=\sum_{\bk\in 2^I}\sum_{\bs\subseteq\bk}\tilde B^\bk_\bs A^\bk_\bs.
\end{equation}
where we have used relation Eq.(\ref{bk}) and the fact $A_\bs^\bk=A_{\bs\wedge\bk}^\bk$ (non-signaling). This new set of coefficients satisfy  $\tilde B_\bs^\bk=\tilde B_{\bs\wedge\bk}^\bk$ and consequently  the coefficients 
\begin{equation}
\tilde B(\ba_\bs,\bs)=\sum_{\bk\in2^I}\frac{(-1)^{\ba_\bs\cdot \bk}}{2^{n-|\bk|}}\tilde B^\bk_{\bs\wedge \bk}
\end{equation}
define the same Bell's inequality as $B(\ba_\bs,\bs)$ does while the second conditions in Eq.(\ref{cd}) are satisfied. \hfill $\square$

Some remarks are in order. In the most general scenario probability and correlation forms of Bell's inequality can be obtained similarly, with coefficients $B(\ba_\bs,\bs)$ satisfying similar conditions as in Eq.(\ref{cd}), which are identical to the non-signaling conditions.  Next we shall consider a special example of Bell's inequality.

\section{Bell-Hardy's inequality}

From the view point of the relation between the entanglement and nonlocality, the Bell-Hardy inequality can be regarded as a most natural generalization to $n$ particles of the well known Bell-CHSH inequality. It was shown by Gisin that every pure entangled two-qubit state violate the Bell-CHSH inequality and recently it has been shown that Bell-Hardy's inequality can be used to established the equivalence between the nonlocality and entanglement for pure states. Bell-Hardy's inequality is based on Hardy's argument or test for nonlocality.  

Let $n$ space-like separated observers, labeled with $I$, perform two dichotomic measurements locally, with setting vector $\bs$ and corresponding outcome vector $\ba_\bs$. Let ${\bf0},{\bf1}$ denote the all zero or one vector and ${\bf 1}_j=[\delta_{jk}]_{k\in I}$. For example a setting vector $\bs={\bf0}$ together with an outcome vector $\ba_\bs={\bf 1}$ means that observer $k$ measures the observable $A_{k|0}$ and obtains 1 as outcome  for each $k\in I$. The probability of this event is denoted as $P({\bf1|\bf 0})$. Hardy's test consists of the following $n+2$ conditions 
\begin{equation}\label{hcd}
P({\bf1}|{\bf0})>0, \quad P({\bf0}|{\bf1})=0, \quad P({\bf1}|{\bf1}_j)=0\quad (j\in I).
\end{equation}
For any local realistic model, where the probabilities assume the local form Eq.(\ref{P}), these conditions cannot be satisfied simultaneously. For example from the first condition it follows that there is a set of hidden variables $\lambda$ with a nonzero measure such that the response functions $a^\lambda_{k|0}=1$ for every observer $k\in I$. For these hidden variables, it follows from last $n$ conditions that $a_{j|1}^\lambda=0$ for all $j\in I$, which contradicts with the second condition.
Hardy's test gives rise to the following  Bell-Hardy inequality
\begin{equation}\label{BH}
 \sum_{j\in I}P({\bf1}|{\bf1}_j)+P({\bf0}|{\bf1})-P({\bf1}|{\bf0})\ge 0.
\end{equation}
Hardy's test was for two particles and  Mermin  formulated Hardy's inequality for two qubits \cite{mermin.hd}, which is equivalent to Bell-CHSH inequality and Cereceda \cite{cereceda} proposed the  $n$-particle version.
By using the notions of Bell polytope, we can rewrite Bell-Hardy inequality simply as $\langle H_n|P\rangle\ge0$ for all $|P\rangle\in \mB_n$ where
\begin{equation}\label{H}
|H_n\rangle=\sum_{j=1}^n|{\bf 1}_j\rangle_I\otimes |{\bf 1}\rangle_I+|{\bf 1}\rangle_I\otimes|{\bf0}\rangle_I-|{\bf 0}\rangle_I\otimes|{\bf 1}\rangle_I.
\end{equation}
If a quantum mechanical state passes Hardy's test, i.e., all conditions in Eq.(\ref{hcd}) are satisfied, then it naturally leads to a violation to Bell-Hardy inequality given above. On the other hand there are entangled pure states that fail Hardy's test while violating the Bell-Hard inequality. To prove that the Bell-Hardy inequality holds for any local realistic model we have only to show that for all extremal local boxes $|\ba;\bb\rangle\in P_n$ the inequality holds 
\begin{eqnarray*}
\langle H_n|\ba;\bb\rangle&=&\sum_{j\in I}b_{j}\prod_{k\not= j}a_{k}+\prod_{k\in I}(1-b_k)-\prod_{k\in I}a_k\\
&\ge&\left(\sum_{j\in I}b_{j}+\prod_{k\in I}(1-b_k)-1\right)\prod_{k\in I}a_k\ge0.
\end{eqnarray*}

{\bf Theorem 1 }Bell-Hardy's inequality Eq.(\ref{BH}) is tight.

{\bf Proof. }We need to show that there are $3^n-1$ independent extremal vectors $|\ba;\bb\rangle\in P_n$, defined in Eq.(\ref{ep}), that saturate Bell-Hardy inequality $\langle H_n|\ba;\bb\rangle=0$. In order to count the dimension $d_H=\dim \mV_n$ of the subspace
$\mV_n=\spn(V_n)$ of all possible linear combinations of vectors in
 \begin{equation}
V_n=\{|{\ba;\bb\rangle}|\langle H_n| \ba;\bb\rangle=0\}
 \end{equation}
the basic idea is to add one more vector that does not belong to $\mV_n$, namely $|{\bf 0;\bf 0}\rangle$, to $V_n$ and prove that $\mP_n=\spn(\tilde V_n)$ with $\tilde V_n=\{|{\bf 0;\bf 0}\rangle\}\cup V_n$, from which it follows $d_H=\dim\mP_n-1$. First of all we have $|{\bf 0;\bf 0}\rangle\not \in \mV_n$ because $\langle H_n|{\bf 0;\bf 0}\rangle=1$ while all vectors in $\mV$ satisfy 
$\langle H_n|V\rangle=0$. Secondly, we observe 
 \begin{equation}\label{U}
\spn(U_n\cup\{|{\bf0;\bf0}\rangle\})=\spn(U_n^\prime\cup\{|{\bf 0;\bf 1}\rangle\})=\mP_n
\end{equation}
where $U_n=\{|\ba;\bb\rangle|\bb\not={\bf 0}\}$ and $U^\prime_{n}=\{|{\ba;\bb}\rangle|\ba\not=\bf 1\}$. To show this, let $u_n$ be the number of independent vectors in $U_n$ and obviously $u_1=2$. Denoting
$\ba=(a_1,\ba^\prime)$ and $\bb=(b_1,\bb^\prime)$ with $\ba^\prime$ and $\bb^\prime$ being two $(n-1)$ dimensional binary vectors, we have a partition $U_n=\cup_{a_1,b_1=0,1}U_{a_1b_1}$ where
\begin{equation*}
U_{a_1b_1}=\{ |{a_1;b_1}\rangle\otimes|{\ba^\prime ;\bb^\prime}\rangle|(b_1,\bb^\prime)\not={\bf 0}\}\quad (a_1,b_1=0,1).
\end{equation*}
It is clear that $U_{a_1,1}= |{a_1;1}\rangle\otimes P_{n-1}$ has $3^{n-1}$ independent vectors for each $a_1=0,1$ while $U_{0,0}=|{0;0}\rangle\otimes U_{n-1}$ has $u_{n-1}$ independent vectors. All the vectors in $U_{1,0}$ can be obtained by linear combinations of vectors in other 3 sets due to identity Eq.(\ref{id}). Thus we have $u_n=u_{n-1}+2\times 3^{n-1}$ which gives $u_n=3^n-1$.
Moreover we have $|{\bf0;\bf0}\rangle\not\in \spn (U_n)$, due to the identity $\langle \ba;\bb|{\bf 1}\rangle_I=\langle\bb|_I$, and thus
  $\spn(U_n\cup\{|{\bf0;\bf0}\rangle\})=\mP_{n}$. The statement for $U^\prime_{n}$ can be proved in the same manner by noting that $|{\bf0;\bf1}\rangle\not\in \spn (U_n^\prime)$.

We shall proceed with induction. Obviously it is true if $n=1$, i.e., $\mP_1=\spn(\tilde V_1)$ with $V_1=\{|1;1\rangle,|1;0\rangle\}$. Suppose $\mP_n=\spn(\tilde V_n)$ is true for $n$.
By denoting $(\ba,a_{n+1})$ and $(\bb,b_{n+1})$ two $(n+1)$ dimensional binary vectors,
it is easy to check the following statements are true
\begin{itemize}
\item[1.] $|(\ba,0);(\bb,0)\rangle=|\ba;\bb\rangle\otimes |0;0\rangle\in V_{n+1}$ if $\bb\not={\bf 0}$, i.e, $U_n\otimes |{0;0}\rangle\subset V_{n+1}$, 
because
\begin{equation*}
\langle H_{n+1}|(\ba,0);(\bb,0)\rangle=\prod_{k\in I}\delta(b_k,0);
\end{equation*}
\item[2.] $|(\ba,0);(\bb,1)\rangle=|\ba;\bb\rangle\otimes |0;1\rangle\in V_{n+1}$ if $\ba\not={\bf 1}$, i.e, $U_n^\prime\otimes |{0;1}\rangle\subset V_{n+1}$, because 
\begin{equation*}
\langle H_{n+1}|(\ba,0);(\bb,1)\rangle=\prod_{k\in I}\delta(a_k,1);
\end{equation*}
\item[3.] $|(\ba,1);(\bb,0)\rangle=|\ba;\bb\rangle\otimes |1;0\rangle\in V_{n+1}$ if $|\ba;\bb\rangle\in V_n$, i.e, $V_n\otimes |{1;0}\rangle\subset V_{n+1}$, because
\begin{equation*}
\langle H_{n+1}|(\ba,1);(\bb,0)\rangle=\langle H_n|\ba;\bb\rangle;
\end{equation*}
\item[4.] $|{\bf0;\bf0}\rangle\otimes |1;1\rangle\in V_{n+1}$;
\item[5.] $|{\bf1;\bf0}\rangle\otimes |1;1\rangle\in V_{n+1}$.
\end{itemize}
By adding one single vector $|{\bf 0;\bf 0}\rangle\otimes|{0;0}\rangle\in\tilde V_{n+1}$ to the subset $U_n\otimes  |{0;0}\rangle$ we see that $\mP_n\otimes |{0;0}\rangle \subset \spn(\tilde V_{n+1})$, taking into account Eq.(\ref{U}). In particular we have
\begin{itemize}
\item[6.] $|{\bf 1;\bf 0}\rangle\otimes |{0;0}\rangle\in \spn(\tilde V_{n+1})$.
\end{itemize}
 Because of observations 2 and 4, together with  identity Eq.(\ref{id}), we have $|{\bf 0;\bf 0}\rangle\otimes |{1;0}\rangle \in  \spn(\tilde V_{n+1})$ so that, due to observation 3, $\mP_n\otimes|{1;0}\rangle=\spn(\tilde V_n)\otimes|{1;0}\rangle$ is also a subset of $\spn(\tilde V_{n+1})$. In particular we have
 \begin{itemize}
 \item[7.]$|{\bf 1;\bf 0}\rangle\otimes|{1;0}\rangle\in\spn(\tilde V_{n+1})$. 
 \end{itemize}
As a result of observations 5,6, and 7, taking into account identity Eq.(\ref{id}), we have $|{\bf 1;\bf 0}\rangle\otimes|{0;1}\rangle\in\spn(\tilde V_{n+1})$ from which it follows $\mP_n\otimes |{0;1}\rangle$ is also a subset of $\spn(\tilde V_{n+1})$, taking into account Eq.(\ref{U}). All these results, together with identity Eq.(\ref{id}),  infer further $\mP_n\otimes|{1;1}\rangle\subset \spn(\tilde V_{n+1})$, from which it follows $\mP_{n+1}=\cup_{x,y=0,1}\mP_n\otimes|{x;y}\rangle\subseteq\spn(\tilde V_{n+1})$. On the other hand we have $\tilde V_{n+1}\subseteq P_{n+1}$, meaning that $\spn(\tilde V_{n+1})\subseteq\mP_{n+1}$, which leads to $\spn(\tilde V_{n+1})=P_{n+1}$ if $\spn(\tilde V_n)=P_n$.\hfill $\square$

Similar methods have been used in \cite{augu} to derive some other tight Bell's inequalities.

\section{Non-signaling polytope}

Quantum mechanical predictions violate Bell's inequality, rendering quantum theory nonlocal. However, quantum theory is not that nonlocal as it obeys the so called non-signaling conditions, i.e., local choices of measurement settings cannot be used to send information. More precisely, in the given $(n,2,2)$ scenario, a set of $4^n$ probabilities $\tilde P(\bb_\bt|\bt)$ are non-signaling if it holds
\begin{equation}
\sum_{b_{k|t_k}=0}^1\tilde P(\bb_\bt|\bt)\quad \mbox{is independent of}\quad t_k \quad (\forall k\in I).
\end{equation}
That is to say, by summing out all the outcomes of the local measurement of a given observer, the joint probability of all other local measurements is independent of the choice of the measurement setting by the given observer. Obviously local boxes are non-signaling. From these non-signaling constraints, together with the normalization of the probability distribution for each given measurement setting, there are $3^n-1$ independent parameters in a general non-signaling joint probability. Conveniently these parameters can be taken as 
\begin{equation}
\tilde A^\bk_\bt=\sum_{\bb_\bt\in 2^I}(-1)^{\bk\cdot\bb_\bt}\tilde P(\bb_\bt|\bt) \quad \mbox{for}\quad \bt\subseteq\bk
\end{equation}
with $\tilde A^\bk_\bt=\tilde A^\bk_{\bt\wedge\bk}$ and for a general $\bt$ reflecting the non-singaling constraints and $\tilde A^{\bf 0}_{\bf 0}=1$ reflecting the normalization. Every set of non-signaling probabilities, referred to as a non-singaling box,  also corresponds to a vector $|\tilde P\rangle$ in the $4^n$ dimensional box space $\mR$. The non-negativeness conditions $\tilde P(\bb_\bt|\bt)\ge 0$ for all possible $4^n$ events define the non-signaling polytope $\mP_{ns}$ with facet given by conditions $\tilde P(\bb_\bt|\bt)= 0$. If a non-signaling box is uniquely determined by its zeros and normalization, then the facets defined by its zeros intersect at a single point, which is a vertex of the non-signaling polytope and it is called as an extremal box.

In certain sense quantum theory lies between the Bell polytope and the non-signaling polytope. On the one hand local realistic models cannot reproduce quantum theory and on the other hand there are non-signaling boxes, e.g., Popescu-Rohrlich box, whose correlations cannot be reproduced by quantum theory. Quite recently a thorough numerical searches for all the extremal tripartite boxes and tight Bell's inequalities has been performed and it turns out that they have the same number \cite{3}. Later on this duality is rigorously established by Fritz \cite{fritz}.
Our second result is the following alternative explicit formula for this duality.

{\bf Theorem 2 }Every Bell's inequality in its standard from as given in Eq.(\ref{bi}) with $B(\ba_\bs,\bs)$ satisfying conditions Eq.(\ref{cd}) defines a non-signaling box
\begin{equation}\label{nb}
\tilde P(\bb_\bt|\bt)=\sum_{\bs\in 2^I}B(\bb_\bt+\bs\wedge \bt,\bs),
\end{equation}
where the addition of two binary vectors is always module 2. Every non-signaling box $\{\tilde P(\bb_\bt|\bt)|\bt,\bb_\bt\in 2^I\}$ defines the following Bell's inequality in its standard form
\begin{equation}\label{nb2}
\sum_{\bs,\bt\in 2^I}\sum_{\ba_\bs,\bb_\bt\in 2^I}\frac{3^n\tilde P(\bb_\bt|\bt)P(\ba_\bs|\bs)}{8^n(-3)^{|\ba_\bs+\bb_\bt+\bs\wedge\bt|}}=\sum_{\bk\in 2^I}\sum_{\bs,\bt\subseteq\bk}\frac{(-1)^{\bs\cdot\bt}}{2^{n+|\bk|}}\tilde A^\bk_\bt A^\bk_\bs\ge0.
\end{equation}
If Bell's inequality Eq.(\ref{bi}) is tight then the corresponding non-signaling box Eq.(\ref{nb}) is extremal and if the nonlocal box $\tilde P(\bb_\bt|\bt)$  is extremal then the corresponding Bell's inequality Eq.(\ref{nb2}) is tight.  

{\bf Proof. } Given an arbitrary Bell's inequality for probabilities as in Eq.(\ref{bi}), it is clear that $\tilde P(\bb_\bt|\bt)$ defined in Eq.(\ref{nb}) is nonnegative, due to  the fact that 
\begin{equation}
\tilde P(\bb_\bt|\bt)=\sum_{\bs,\ba_\bs\in 2^I}B(\ba_\bs,\bs)\delta(\ba_\bs,\bb_\bt+\bs\wedge\bt)=\langle B| \bb_\bt;\bb_\bt+\bt\rangle
\end{equation}
and $| \bb_\bt;\bb_\bt+\bt\rangle\in P_n$. Here we have used the fact
$$\sum_{\bs,\ba_\bs}\delta(\ba_\bs,\bb_\bt+\bs\wedge\bt)|\bs;\ba_\bs\rangle
=\bigotimes_{k\in I}\Big(|0\rangle_k\otimes|b_{k|t_k}\rangle_k+|1\rangle_k\otimes|b_{k|t_k}+t_k\rangle\Big)=| \bb_\bt;\bb_\bt+\bt\rangle.
$$
From conditions Eq.(\ref{cd}) it follows that Eq.(\ref{nb}) is a normalized probability distribution, since $\langle B|\theta\rangle=1$, satisfying the non-signaling conditions, i.e., $\tilde P(\bb_\bt|\bt)$ is a non-signaling box. If the given Bell's inequality is tight, i.e., there are $3^n-1$ independent extremal vectors such that $\langle B|\ba;\bb\rangle=0$ and we denote 
\begin{equation*}
V_B=\{|\ba;\bb\rangle\mid\langle B|\ba;\bb\rangle=0\}.
\end{equation*}
For each extremal vector $|\ba;\bb\rangle\in V_B$ we obtain a zero of the non-signaling box $\tilde P(\bb_\bt|\bt)$ by choosing the measurement settings according to $\bt=\bb+\ba$ with outcomes given by $\bb_\bt=\ba$, i.e., $\tilde P(\ba|\ba+\bb)=0$. We note there exists a one-to-one correspondence
\begin{equation}
\left\{\begin{array}{l}\ba=\bb_\bt\\\bb=\bb_\bt+\bt\end{array}\right.,\quad
\left\{\begin{array}{l}\bt=\ba+\bb\\\bb_\bt=\ba\end{array}\right..
\end{equation}
Because $3^n-1$ equations $\langle B|\ba;\bb\rangle=0$ plus the normalization $\langle B|\theta\rangle=1$ determine the vector $|B\rangle$ uniquely, since $|\theta\rangle\not\in\spn(V_B)$, the corresponding $3^n-1$ zeros of $\tilde P(\bb_\bt|\bt)$ plus the normalization determine the non-signaling box uniquely, meaning that the non-siginaling box is extremal, i.e., a vertex of the non-signaling polytope.

On the other hand, for a given non-signaling box $\tilde P(\bb_\bt|\bt)$ if we can solve from Eq.(\ref{nb}) some function $B(\ba_\bs,\bs)$, then this function will define a Bell's inequality via Eq.(\ref{bi}) since  $\langle B|\ba;\bb\rangle=\tilde P(\ba|\ba+\bb)\ge0$  for all possible extremal vectors $|\ba;\bb\rangle\in P_n$ of the Bell polytope. By denoting
\begin{equation*}
 B_\bs^\bk=\frac1{2^{|\bk|}}\sum_{a_\bs\in 2^I}(-1)^{\ba_\bs\cdot \bk}B(\ba_\bs,\bs),\quad
 B(\ba_\bs,\bs)=\sum_{\bk\in2^I}\frac{(-1)^{\ba_\bs\cdot \bk}}{2^{n-|\bk|}}B^\bk_{\bs}
\end{equation*}
we obtain 
\begin{eqnarray*}
\tilde A^\bk_\bt
&=&\sum_{\bb_\bt\in 2^I}(-1)^{\bk\cdot\bb_\bt}\tilde P(\bb_\bt|\bt) \\
&=&\sum_{\bb_\bt\in 2^I}(-1)^{\bk\cdot\bb_\bt}\sum_{\bs\in 2^I}B(\bb_\bt+\bs\wedge \bt,\bs) \\
&=&\sum_{\bb_\bt\in 2^I}(-1)^{\bk\cdot\bb_\bt}\sum_{\bs\in 2^I}\sum_{\bk^\prime\in2^I}\frac{(-1)^{(\bb_\bt+\bs\wedge \bt)\cdot \bk^\prime}}{2^{n-|\bk|}}B^{\bk^\prime}_{\bs}\\
&=&2^{|\bk|}\sum_{\bs\in 2^I}(-1)^{(\bs\wedge\bk)\cdot \bt}B^{\bk}_{\bs}\\
&=&2^n\sum_{\bs\subseteq\bk}(-1)^{\bs\cdot\bt}B^\bk_\bs
\end{eqnarray*}
As a result we have 
\begin{equation}
 B^\bk_\bs=\frac1{2^{n+|\bk|}}\sum_{\bt\subseteq\bk}(-1)^{\bs\cdot\bt}\tilde A^\bk_\bt,
\end{equation}
from which the Bell's inequality for correlations follows, and 
\begin{eqnarray*}
B(\ba_\bs,\bs)&=&
\sum_{\bk,\bt,\bb_\bt\in 2^I}\frac{(-1)^{(\ba_\bs+\bb_\bt+\bs\wedge\bt)\cdot\bk}}{2^{3n-|\bk|}}\tilde P(\bb_\bt|\bt)\\
&=&
\sum_{\bt,\bb_\bt\in 2^I}\frac{\tilde P(\bb_\bt|\bt)}{8^n}\prod_{k\in I}\left(1+2(-1)^{a_{k|s_k}+b_{k|t_k}+s_kt_k}\right)\\
&=&\sum_{\bt,\bb_\bt\in 2^I}\frac{3^n\tilde P(\bb_\bt|\bt)}{8^n(-3)^{|\ba_\bs+\bb_\bt+\bs\wedge\bt|}},
\end{eqnarray*}
from which the Bell's inequality for probabilities  follows.
Again, for each zero of the non-signaling box $\tilde P(\bb_\bt|\bt)=0$ there is an extremal vector $|\bb_\bt;\bb_\bt+\bt\rangle\in P_n$ of Bell polytope such that $\langle B|\bb_\bt;\bb_\bt+\bt\rangle=0$, i.e., $|\bb_\bt;\bb_\bt+\bt\rangle\in V_B$. If the non-signaling box is extremal, i.e., uniquely determined by its zeros and normalization, then there are no less than $3^n-1$ extremal vectors of the Bell polytope in $V_B$. The conditions $\langle B|\ba;\bb\rangle=0$ with $|\ba;\bb\rangle \in V_B$, together with the normalization $\langle B|\theta\rangle=1$,  should determine the non-signaling box and thus $|B\rangle$ uniquely, meaning that there are at least $3^n-1$ independent  extremal vectors in $V_B$, i.e., the corresponding Bell's inequality is tight. \hfill $\square$

As an example, let us consider the Bell-Hardy inequality Eq.(\ref{BH}) or equivalently $\langle H_n|P\rangle\ge0$ with $|H_n\rangle$ given in Eq.(\ref{H}). This form of Bell-Hardy inequality is not standard because the coefficients 
\begin{equation}
\langle H_n|\bs\rangle_I\otimes|\ba_\bs\rangle_I=\sum_{j\in I}\delta_{\bs,{\bf1}_j}\delta_{\ba_\bs,{\bf 1}}+\delta_{\bs,{\bf 1}}\delta_{\ba_\bs,{\bf 0}}-\delta_{\bs,{\bf 0}}\delta_{\ba_\bs,{\bf 1}}
\end{equation}
 do not satisfy the conditions in Eq.(\ref{cd}). First of all we have $\langle H_n|\theta\rangle=n$ and by denoting
 \begin{equation*}
 H_\bs^\bk=\sum_{\ba_\bs\in 2^I}(-1)^{\bk\cdot\ba_\bs}\langle H_n|\bs\rangle_I\otimes|\ba_\bs\rangle_I=\sum_{j\in I}(-1)^{|\bk|}\delta_{\bs,{\bf1}_j}+\delta_{\bs,{\bf 1}}-(-1)^{|\bk|}\delta_{\bs,{\bf 0}}
 \end{equation*}  
the Bell-Hardy inequality can be cast into the following correlation form 
\begin{eqnarray*}
0\le\langle H_n|P\rangle&=&\frac1{2^n}\sum_{\bk,\bs\in2^I}H_\bs^\bk A_\bs^\bk\\&=&\frac1{2^n}\sum_{\bk\in2^I}\sum_{\bs\subseteq\bk}A_\bs^\bk\sum_{\bt\in 2^I}
H_\bt^\bk \delta_{\bt\wedge\bk,\bs\wedge\bk}
:=n\sum_{\bk\in2^I}\sum_{\bs\subseteq\bk}A_\bs^\bk\tilde H_\bs^\bk
\end{eqnarray*}
where
\begin{equation}
\tilde H^\bk_\bs=\frac1{n2^{n}}\left((-1)^{|\bk|}\sum_{j\in I}\delta_{\bs,{\bf1}_j\wedge\bk}+\delta_{\bs,{\bf 1}\wedge\bk}-(-1)^{|\bk|}\delta_{\bs,{\bf 0}\wedge\bk}\right)
\end{equation} 
so that the following coefficients 
 \begin{equation}
 \tilde H(\ba_\bs,\bs)=\sum_{\bk\in 2^I}\frac{(-1)^{\bk\cdot\ba_\bs}}{2^{n-|\bk|}}\tilde H^\bk_{\bs\wedge\bk}
 \end{equation}
 give rise to the standard probability form of Bell-Hardy inequality.
Thus the extremal non-signaling box corresponding to the Bell-Hardy inequality via Theorem 2 has the following correlations
\begin{equation}
\tilde A^\bk_\bt=2^n\sum_{\bs\subseteq\bk}(-1)^{\bs\cdot\bt}\tilde H_\bs^\bk=\frac{\sum_{j\in I}(-1)^{\bt_j\cdot\bk}+(-1)^{\bt\cdot\bk}-(-1)^{|\bk|}}n
\end{equation}
where  $\bt_j={\bf1}+\bt\wedge{\bf 1}_j=[1-t_j\delta_{kj}]_{k\in I}$ for each $j\in I$, 
from which we obtain finally the corresponding extremal non-signaling box, referred to as Hardy box here,
\begin{equation}
\tilde P_H(\bb_\bt|\bt)=\frac{\sum_{j\in I}\delta_{\bb_\bt,\bt_j}+\delta_{\bb_\bt,\bt}-\delta_{\bb_\bt,\bf 1}}n.
\end{equation}
In the case of $n=3$  the Hardy box is the extreme box number 29 listed in \cite{3}.
We note that the single particle probability of obtaining outcome 0 or 1 is $1/n$ or $(n-1)/n$ for both two choices of local measurements. Moreover this extremal non-signaling box can be easily proved to be nonlocal, i.e., violating some Bell's inequalities, e.g., the Bell inequality due to Werner \& Wolf \cite{ww} and Zukowski \& Brukner \cite{zb}.

\section{Conclusions and discussions}

We have shown that the Bell-Hardy inequality arising form Hard's nonlocality test is tight and derive an extremal nonlocal box via an explicit duality between the Bell polytope and the non-signaling polytope in the $(n,2,2)$ scenario due to Fritz. Most recently Hardy's test is generalized to the detection of  genuine multipartite nonlocality \cite{chen.q,yu.oh}. It has been shown that fully entangled symmetric pure states of $n$ qubits as well as  all fully entangled pure states  are genuinely multipartite nonlocal. That is to say, to reproduce quantum correlations in a fully entangled symmetric state or a three particles pure states, those hybrid local/nonlocal realistic models have to be signaling. We believe that the extremal non-signaling box defined here is also genuinely multipartite nonlocal. Finally,
as a possible generalization, the explicit duality between Bell polytope and non-signaling polytope worked out here may lead to a partial duality in other scenarios.   

This work is funded by the Singapore Ministry of Education (partly through the Academic Research Fund Tier 3 MOE2012-T3-1-009).

\end{document}